\documentclass[11pt]{article}
\usepackage{a4wide}
\usepackage{authblk}
\usepackage{graphicx}
\usepackage{caption}
\usepackage{subcaption}
\usepackage{color}
\usepackage{amsmath,amsfonts,amssymb,graphicx,hyperref,hypcap}
\usepackage{epsfig}
\usepackage{amsmath,amsfonts,epsf}
\usepackage{amssymb,tikz}

\newcommand{\done}[2]{\frac{d{#1}}{d{#2}}}

\begin{document}
\begin{center}
{\Large\bf 
Realising Interactions Between Dark Matter and Dark Energy Using $k$-essence Cosmology}
\end{center}
\vspace{4mm}
\begin{center}
{\large Abhijit Bandyopadhyay\footnote{Email: abhijit@rkmvu.ac.in} and Anirban Chatterjee\footnote{Email: anirban.chatterjee@rkmvu.ac.in}}
\end{center}

\begin{center}
Department of Physics\\Ramakrishna Mission Vivekananda University\\
Belur Math, Howrah 711202, India
\end{center}
\vspace{4mm}

\begin{abstract}
In this paper we exploit dynamics of a $k-$essence scalar field  to realise 
interactions between dark components of universe
resulting in a evolution consistent with observed features of late time phase of cosmic evolution. 
Stress energy tensor corresponding to a $k-$essence Lagrangian $L=V( \phi)F(X)$ (where $X=\frac{1}{2}g^{\mu\nu}\nabla_\mu\phi
\nabla_\nu\phi$) is
shown to be equivalent to an ideal fluid with two components having same equation of state. 
Stress energy tensor of one of the components
may be generated from a constant potential $k-$essence Lagrangian of form $L_1=V_0F(X)$ ($V_0$ constant)
and that of other from another Lagrangian of form $L_2=V_1(\phi)F(X)$ with $V=V_0 + V_1(\phi)$.
 We have shown that, the unified dynamics of dark matter and dark energy
described by a single scalar field $\phi$ driven by a 
$k-$essence Lagrangian $L= V(\phi)F(X)$
may be viewed in terms of diffusive interactions between the two 
hypothetical fluid components  `1' and `2' with
stress energy tensors equivalent to that of Lagrangians $L_1$ and $L_2$ respectively.
The energy transfer between the fluid components
is determined by functions $V(\phi)$, $F(X)$ and their derivatives.
Such a realisation is shown to be consistent with
the Supernova Ia data with certain constraints on the 
temporal behaviour of $k-$essence potential $V(\phi)$.
We have described a methodology to obtain such constraints.

\end{abstract}

\section{Introduction}	
\label{sec:1}
Till date there exists strong experimental evidence  in support of following
facts related to late time phase of cosmic evolution:
\begin{enumerate}
\item[(a)] At large scales, universe appears isotropic and homogeneous to comoving observers. \cite{wu1}

\item[(b)] The universe has undergone a transition from a phase of decelerated expansion
to  accelerated expansion during its late time phase of evolution. 
Observations of red-shifts and luminosity distances of 
~type Ia Supernovae (SNe Ia),   \cite{Riess1}
~Baryon Acoustic Oscillations,  \cite{ref:Cole}  
provide overwhelming 
evidences in favour  of this fact. Source of this late-time
cosmic acceleration is generally labeled as `Dark Energy' (DE).

\item[(c)] Besides baryonic matter, 
there exists non-luminous matter in present universe,
indirectly
manifesting its existence through gravitational interactions
as revealed in observation of rotation curves of
spiral ~galaxies,\cite{Sofue:2000jx} ~gravitational 
lensing,\cite{Bartelmann:1999yn} ~Bullet cluster,\cite{Clowe:2003tk}
and other colliding  clusters. Such `matter' is termed as `Dark Matter' (DM).

\item[(d)] At present epoch, dark content (DE and DM) of universe
contributes 96\% ($\sim 70\%$ dark energy and $\sim 26\%$ dark matter)
of total energy density of the universe. Rest $\sim 4\%$ is contributed by 
baryonic matter with negligible contribution from radiations. 
This has been established by
measurements in satellite borne experiments - WMAP, \cite{Hinshaw:2012aka} and Planck  \cite{Ade:2013zuv}.
\end{enumerate}
The origin of onset of late time cosmic acceleration
(dark energy) still remains a mystery.
Diverse theoretical models of dark energy have been
constructed to explain dark energy.
There exist field theoretic models of dark energy
involving consideration of specific forms of the energy-momentum
tensor in Einstein's equation. They include 
quintessence models, \cite{ref:quint} in which potential energy of a scalar
field drives the cosmic acceleration and 
$k-$essence models \cite{ref:kess}  where accelerated expansion
arises due to modifications to kinetic energy
of scalar fields.
Other viable dark energy models, constructed by modifying geometric part of Einstein's equation,
include $f(R)$-gravity \cite{fr1}, 
scalar-tensor theories \cite{st1} and
brane world models \cite{brm1}.\\

A physical theory for origin and nature of dark matter and dark energy is yet lacking.
However, $\Lambda$CDM model \cite{ref:weinberg89, Bamba} provides a parametrization of the model of cosmic evolution
that fits a wide variety of 
cosmological data. The principal ingredients of $\Lambda$CDM model
are cold dark matter (CDM) and Cosmological constant $(\Lambda)$ 
representing  dark energy of vacuum. 
But the model is plagued with two major problems: ($i$) fine tuning problem
- large disagreement between vacuum expectation value of energy
momentum tensor  and  observed value of dark
energy density and ($ii$) the coincidence
problem -  order of magnitude of observable values of dark matter and dark energy  at present epoch
are same. 
There exist diverse theoretical
approaches aiming to address above mentioned issues. One such
interesting approach, which specifically addresses the
issue of coincidence problem, involves consideration of a unified model of
dark energy and dark matter
with a diffusive energy transfer between dark energy and dark matter.
Different aspects of dynamics of this model are discussed in detail 
in \cite{Szydlowski, calogero, calogero1, calogero2, Haba:2016swv,Ber1,Ber2}. 
 In the framework of such models of dark fluids involving interactions
between dark energy and dark matter, 
the dark matter and dark energy densities
at any epoch are correlated.
However, to seek for any quantitative explanation for coincidence
of dark matter and dark energy densities at present epoch
one needs to construct microscopic models of dark fluids
in a way that their dynamics would give rise to
correlations between dark matter and dark energy densities
at any epoch.\\

The basic framework for the model of diffusive DM-DE interaction
is as follows. 
The total stress energy tensor $T_{\mu\nu}$
appearing in right hand side of
Einstein field equations
$
R_{\mu\nu} - (1/2)g_{\mu\nu}R = (8\pi G/c^4)T_{\mu\nu}$
is conserved \textit{ i.e.} $\nabla^{\mu} T_{\mu\nu} = 0$. 
($R_{\mu\nu}$ is the Ricci tensor, 
$g_{\mu\nu}$ is the metric,  $G$ and $c$ are  Newton's constant
and  velocity of light respectively).
The total energy-momentum tensor $T_{\mu\nu}$  
may be decomposed into contributions from radiation ($T_R^{\mu\nu}$), baryonic matter ($T_b^{\mu\nu}$), dark matter
($T_{\rm dm}^{\mu\nu}$) 
and dark energy $T_{\rm de}^{\mu\nu}$ as
$T_{\mu\nu} = T^R_{\mu\nu} + T^b_{\mu\nu} + T^{\rm dm}_{\mu\nu} + T^{\rm de}_{\mu\nu}$.
 Based on observations from  
 WMAP and PLANCK experiments, as discussed in Sec.\ \ref{sec:1},
we neglect  the contribution of baryonic matter and radiation
to the total energy density of the universe  during late time phase of cosmic evolution, which is the relevant  domain of cosmic time probed in 
SNe Ia observations. Thus for late time phase of
cosmic evolution the conservation equation  $\nabla^{\mu} T_{\mu\nu} = 0$
implies $\nabla_\mu(T_{dm}^{\mu\nu} + T_{de}^{\mu\nu}) \approx 0$
.
An energy transfer between
dark matter and dark energy respecting conservation of total stress tensor
implies 
\begin{eqnarray}
\nabla_\mu T_{de}^{\mu\nu} = - \nabla_\mu T_{dm}^{\mu\nu} \equiv - J^\mu
\label{eq:in1}
\end{eqnarray}
where $J^\mu$ is some current source corresponding to the non-conserved stress tensor $T_{de}^{\mu\nu}$
and $T_{dm}^{\mu\nu}$.  A diffusion in the dark fluid environment is assumed to cause
transfer of energy from dark energy to dark matter in this model as manifested through the 
non-conservation  Eq. \eqref{eq:in1}.\\

In this paper we exploit dynamics of a $k-$essence scalar field $\phi$ to realise 
interactions between dark components of universe
resulting in a evolution consistent with observed features of late time phase of cosmic evolution 
as probed in SNe Ia observations \cite{Riess1}.
$k-$essence scaler field models involve Lagrangian with non-canonical
kinetic terms \cite{ref:kess} expressed as 
\begin{eqnarray}
L = V(\phi)F(X)\,, \label{eq:lag}
\end{eqnarray}
where the kinetic term $X = \frac{1}{2}g^{\mu\nu}\nabla_\mu\phi \nabla_\nu\phi$, $g^{\mu\nu}$ is the metric,
$V(\phi)$ and $F(X)$  are functions of $\phi$ and $X$ respectively. 
The corresponding stress-energy tensor is equivalent to that of
an ideal fluid with energy density $\rho = V(\phi)(2XF_X - F)$ and pressure $p= V(\phi)F(X)$.
We   write the $k-$essence potential  as $V(\phi)=V_0 + V_1(\phi)$, where 
$V_0$ is the constant term in the Taylor expansion of $V(\phi)$ about $\phi=0$
and $V_1(\phi)$ is the $\phi$-dependent part with $V_1(\phi)=0 $  for $\phi=0$.
Using this, $\rho$ and $p$ may be decomposed
as $\rho=\rho_1+\rho_2$ and $p=p_1+p_2$, with $\rho_1=V_0(2XF_X - F)$, $p_1=V_0 F(X)$
and $\rho_2=V_1(\phi)(2XF_X - F)$, $p_2=V_1(\phi) F(X)$ where $F_X = dF/dX$.
Thus the ideal fluid characterised by ($\rho$,$p$) is equivalent
to a 2-component fluid - the components begin labelled by `1' and `2'.
The two components  `1' and `2' are also ideal fluids
characterised by their respective energy densities and pressures as
($\rho_1$,$p_1$) and ($\rho_2$,$p_2$).
Energy momentum tensor of fluids 1 is equivalent to 
that of a $k-$essence Lagrangian with constant potential $L_1 = V_0 F(X)$ and
that of fluid 2 is equivalent to that of a   
$k-$essence Lagrangian   $L_2 = V_1(\phi) F(X)$
where the potential $V_1(\phi)$ is such a function of $\phi$ that,
at $\phi=0$, $V_1(\phi)=0$. In this sense, the two interacting fluids 1 and 2 are different
however though they have same equation of state as $p_1/\rho_1 = p_2/\rho_2$. \\

The motivation of the work is as follows.
Dynamics of dark matter and dark energy 
described by a single $k-$essence scalar field
may be shown to be  equivalent to dynamics of
two interacting fluids, one described by
a $k-$essence Lagrangian with constant potential
and the other with a $k-$essence Lagrangian with 
potential having no constant term in its Taylor expansion.
Such a realisation is shown to be consistent with the Supernova Ia
data with certain constraints on the temporal behaviour of 
the original $k−$essence potential $V(\phi)$.
We have described a methodology to obtain such constraints.\\

We have explicitly shown in  Sec.~\ref{sec:2} that, the unified dynamics of dark matter and dark energy
described by a single scalar field $\phi$ driven by a 
$k-$essence Lagrangian $L= V(\phi)F(X)$
may be viewed in terms of diffusive interactions between the two 
hypothetical fluid components  `1' and `2' with
stress energy tensors equivalent to that of Lagrangians $L_1$ and $L_2$ respectively. 
The energy transfer between these fluid components 
is determined by functions $V(\phi)$, $F(X)$ and their derivatives.
In  Sec.\ \ref{sec:3} have discussed the methodology of analysis of observational data
and obtained  temporal behaviour of
equation of state of the   dark fluid of universe
from SNe Ia data.  
In Sec.\  \ref{sec:4} we have shown
how one may establish a connection between 
the energy densities of the hypothetical fluids,
the   $k-$essence potential $V(\phi)$
and the observed quantities like equation of state of the dark fluid.
 We have shown that such a connection may be realised
  in terms of an equation (Eq.\ \ref{eq:recur1}) relating parameters depicting 
temporal behaviour of the potential $V(\phi(t))$ 
and value of a parameter $\alpha_0$, where
$\alpha_0 / (1 - \alpha_0)$ is the value of the ratio $\rho_1/\rho_2$ at present epoch.
In  Sec.\ \ref{sec:5} we have shown how the above relation may be exploited
to constrain nature of temporal behaviour of $k-$essence potential from observational data.
The obtained constraints thus correspond to only those class of $k-$essence potentials $V(\phi)$
for which dynamics of $k-$essence scalar field $\phi$ with Lagrangian $L=V(\phi) F(X)$
may be used to realise DM-DE interactions in terms of 
diffusive interaction between two ideal fluids, both having equations of 
state same as that of total dark sector of universe. The conclusions are presented in
Sec.\ \ref{sec:6}.

\section{Realisation of diffusive DM-DE interactions using dynamics of $k-$essence scalar field}
\label{sec:2}
In this section we set up the  representative equations
for late time cosmic evolution governed by its interactive
dark matter and dark energy contents. We show how
such an evolution may be realised in terms of dynamics 
of a $k-$essence scalar field. 
We model the dark energy and dark matter contents of universe
as ideal fluids with each component being characterised
by their respective energy density and pressure: $(\rho_{\rm de}, p_{\rm de})$
for dark energy and $(\rho_{\rm dm}, p_{\rm dm})$ for dark matter.
In a homogeneous and isotropic spacetime described by Friedman - Lemaitre - Robertson - Walker (FLRW) metric,
the equations governing dynamics of late time cosmic evolution are the following two independent Friedmann equations
\begin{eqnarray}
&& H^2 = \frac{8\pi G}{3} (\rho_{\rm de} + \rho_{dm})  \label{eq:1}\\
&& \frac{\ddot{a}}{a} = - \frac{4\pi G}{3}\left[(\rho_{\rm dm} + \rho_{\rm de}) + 3p_{\rm de} \right] \label{eq:2}
\end{eqnarray}
where $a(t)$ is the FLRW scale factor, $H = \dot{a}/a$.
We take dark matter to be pressure less (dust),  $p_{\rm dm}=0$.
Equation of state of total dark fluid may then
be expressed in terms of scale factor and its 
derivatives from above two equations as
\begin{eqnarray}
\omega 
= \frac{p_{\rm de}}{\rho_{\rm de} + \rho_{\rm dm}}
= -\frac{2}{3}\frac{\ddot{a}a}{\dot{a}^2} - \frac{1}{3}
\label{eq:ome1}
\end{eqnarray}
We have considered a flat spacetime (zero curvature constant)
and neglect contributions from radiation and baryonic matter during late time phase of cosmic evolution. 
From Eqs. \eqref{eq:1} and \eqref{eq:2} we obtain the continuity 
equation for the total dark fluid content of the universe as
\begin{eqnarray}
(\dot{\rho}_{\rm dm} + \dot{\rho}_{\rm de}) + 3 H \left[(\rho_{\rm dm} + \rho_{\rm de}) + p_{\rm de}\right] &=& 0\,.
\label{eq:3}
\end{eqnarray}
Eq. \eqref{eq:3} represents conservation of energy of total dark sector of
universe comprising dark energy and dark matter. 
In DE-DM interaction scenario,
energy transfer between DE and DM  respecting conservation of total stress tensor for the dark sector
may be realised by writing \eqref{eq:3}  as
\begin{eqnarray}
\dot{\rho}_{\rm de}+3H(\rho_{\rm de}+p_{\rm de}) = - \left[\dot{\rho}_{\rm dm}+3H\rho_{\rm dm}\right] \neq 0 \label{eq:4}
\end{eqnarray}
The above  non-conservation equations for 
individual dark components with   non-zero source term correspond to energy transfer between
dark energy and dark matter in the fluid environment.\\

To realise the dynamics of late time cosmic evolution 
governed by its interacting dark matter and dark energy contents 
in terms of a $k-$essence scalar field we identify the total 
stress energy tensor  $(T_{dm}^{\mu\nu} + T_{de}^{\mu\nu})$
of the entire dark sector to that the $k-$essence scalar field $\phi$
with Lagrangian given by Eq.\ (\ref{eq:lag}), so that total energy density
and pressure of the dark sector may be written as
\begin{eqnarray}
p_{de} &=& V(\phi) F(X) \,, \label{eq:6}\\
\rho_{de} + \rho_{dm}  &=& V(\phi) G(X)\, .\label{eq:7}
\end{eqnarray}
where $G(X) = 2XF_X - F$, $ F_X \equiv dF/dX $. This gives
the equation of state of the entire dark fluid in terms
of the function $F(X)$ and its derivatives as
\begin{eqnarray}
\omega 
= \frac{p_{\rm de}}{\rho_{\rm de} + \rho_{\rm dm}} = \frac{F(X)}{G(X)}
\label{eq:omega}
\end{eqnarray}
We assume the scalar field in FLRW background to be homogeneous \textit{i.e.} $ \phi(t,\vec{x}) = \phi(t)$
so that $X = \frac{1}{2}\nabla_\mu\phi\nabla^\mu \phi = \frac{1}{2}\dot{\phi}^2$.
Using  Eqs.\eqref{eq:6} and \eqref{eq:7}, the 
continuity equation for the dark fluid (Eq.\ \eqref{eq:3})
may be put in a form
\begin{eqnarray}
\dot{G} + 3H (G + F) &=& - G\frac{\dot{V}}{V}
\label{eq:eomfg}
\end{eqnarray}
where $\dot{G}$ and $\dot{V}$ are respective time derivatives of $G$ and $V$
having implicit time dependences through $X=(1/2)\dot{\phi}^2(t)$ and
$\phi(t)$ respectively. \\

As already mentioned in Sec.\ \ref{sec:1},
we write $V(\phi) = V_0 + V_1(\phi)$ by separating out a constant part $V_0$ and
$V_1(\phi)$ is a function of $\phi$ such that at $\phi=0$, $V_1(\phi) = 0$.
We construct a 2-component fluid model to understand DM-DE interactions in terms of $k-$essence field $\phi$ 
in the following way.
We consider an ideal fluid in FLRW background with two interacting ideal fluid components
labeled as `fluid 1' and `fluid 2' characterised by their respective densities and pressures 
$(\rho_1,p_1)$ and $(\rho_2,p_2)$ identified as
\begin{eqnarray}
\rho_1 = V_0 G(X) \quad &,& \quad \rho_2 = V_1 (\phi) G(X)\,, \nonumber \\
p_1 = V_0 F(X) \quad &,& \quad p_2 = V_1(\phi)  F(X)
\label{eq:fluid12}
\end{eqnarray}
By this construction Eq. \eqref{eq:fluid12} we have
\begin{eqnarray}
p_1 + p_2 &=& V(\phi) F(X) = p_{\rm de} \nonumber\\
\rho_1 + \rho_2 &=& V(\phi) G(X) = \rho_{\rm de} + \rho_{\rm dm} 
\label{eq:12dedm}
\end{eqnarray}
and equation of states of both fluids `1' and `2' are same as that of the
total dark fluid of the universe 
\begin{eqnarray}
\omega = \frac{p_{\rm de}}{\rho_{\rm de} + \rho_{\rm dm}} = \frac{F}{G} = \frac{p_1}{\rho_1} = \frac{p_2}{\rho_2}  
\label{eq:12eos}
\end{eqnarray}
as evident from set of Eqs. \eqref{eq:fluid12} and \eqref{eq:12dedm}. 
Also note that the stress energy tensor of fluid 1
is equivalent to that a $k-$essence Lagrangian with constant potential 
$L_1 = V_0 F(X)$ and that of fluid 2 
is to a  $k-$essence Lagrangian $L_2=V_1(\phi) F(X)$.
Using Eqs. \eqref{eq:3}, \eqref{eq:eomfg} and \eqref{eq:fluid12}, the continuity equation 
$\frac{d}{dt}({\rho}_{\rm dm} + {\rho}_{\rm de}) + 3 H \left[(\rho_{\rm dm} + \rho_{\rm de}) + p_{\rm de}\right] = 0$
for the entire dark fluid may be expressed in terms of pressures and densities characterising
fluid 1 and 2 as
\begin{eqnarray}
\dot{\rho}_1 + 3H(p_1 + \rho_1) = - \left[\dot{\rho}_2 + 3H(p_2 + \rho_2)\right] = - G\frac{V_0\dot{V}}{V}
\label{eq:12int}
\end{eqnarray}
 Eq. \eqref{eq:12int}  represents 
non-conservation equations for 
individual fluids 1 and 2 with a source term $ G\frac{V_0\dot{V}}{V}$ corresponding
to energy transfer between the fluid components 1 and 2 caused by a diffusion in the fluid environment.\\


Note that, 
we use a single scalar field $\phi$ driven by a 
$k-$essence Lagrangian $L= V(\phi)F(X)$
to describe the unified dynamics of dark matter and dark energy.
Set of Eqs. \eqref{eq:fluid12}, \eqref{eq:12dedm}, \eqref{eq:12eos} and \eqref{eq:12int} enable  us to
view this unified dynamics  
in terms of diffusive interaction  between the two 
hypothetical fluid components labeled as `1' and `2'.
Total energy density ($\rho_{\rm de} + \rho_{\rm dm}$)
and pressure ($p_{\rm de}$) of the dark fluid respectively are equal to the
total energy density $(\rho_1 + \rho_2)$ and pressure $(p_1+p_2)$ of the
fluid comprising  components `1' and `2'.
The realisation of interaction between dark matter and dark energy components  
 in terms of interactions of fluid `1' and `2'
expressed through non-conservation  Eq.\ \eqref{eq:12int} with the source term   
$G(X)V_0\dot{V}(\phi)/V(\phi)$. This term accounts for 
  the energy transfer between these two fluids
due to diffusion and is
determined by functions $V(\phi)$ and $F(X)$ (and their derivatives)
that appear in description of a $k-$essence model.

\section{Behaviour of equation of state of dark fluid from observation}
\label{sec:3}
\begin{figure}[t]
\begin{center}
\includegraphics[scale=.6]{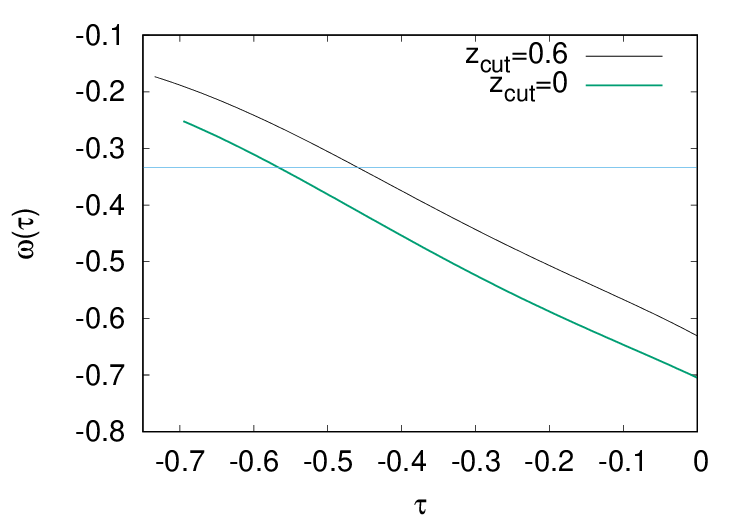} 
\end{center}
\caption{\label{fig:1}Plot of $\omega(\tau)$ vs $\tau$ as obtained from analysis of observational data
for $z_{\rm cut} = 0$ and $z_{\rm cut} = 0.6$ . 
}
\end{figure}

We extract the temporal behaviour of 
 equation of state $\omega(t)$ of the total dark
fluid of the universe from the analysis of SNe Ia observational
data. We later used the observed dependence of $\omega(t)$
to find observational constraints on some parameters 
relevant in the context of  the above mentioned model
of DM-DE interaction.
Here we briefly outline how we analysed 
SNe IA data to extract time dependence of equation
of state from observational data.
The general methodology of analysis of SNe Ia data
has been discussed in detail in \cite{wang:jcap,acab}.
In the context of the present work we have used
time dependence of scale factor
$a(t)$ as an observational input which may be directly
determined from measurement of Luminosity distance and redshifts  
of type Ia Supernova events  within its accessible time domain  
\textit{viz.} $0.4 < t/t_0 <1$ ($t_0$ being the age of the universe).
We discuss below the methodology adopted\cite{wang:jcap} to 
find the temporal behaviour of the scale factor $a(t)$
from the observational data.
A red-shift cut-off ($z_{\rm cut}$) is used there
to separate out SN samples with $z < z_{\rm cut}$
and $z \geq z_{\rm cut}$. For samples with $z < z_{\rm cut}$
the $\chi^2$ has been computed
and for samples with $z \geq z_{\rm cut}$ flux averaged
values of $\mu$ and covariant matrix are used to
compute $\chi^2$ \cite{wang:jcap,acab}.
For our work we take as input, the $z-$dependence of the function $E(z) = H(z)/H_0$,
obtained from the marginalisation of $\chi^2$ shown in left panel 
of Fig.\ 5 of Ref.\ \cite{wang:jcap}. We take $E(z)$ vs $z$ curve
obtained in  \cite{wang:jcap} for two benchmark cases: 
$z_{\rm cut}=0$ and $z_{\rm cut}=0.6$.
We may use the relations $H=\dot{a}/a$ and 
$a_0/a = 1+z$ to write
\begin{eqnarray}
dt &=& - \frac{dz}{(1+z)H(z)} = - \frac{dz}{(1+z)H_0E(z)}
\label{eq:ap1}
\end{eqnarray}
where $E(z) = H(z)/H_0$. The above equation on integration gives
\begin{eqnarray}
\frac{t(z)}{t_0} &=& 1 
- \frac{1}{H_0t_0}\int_z^0 \frac{dz^\prime}{(1+z^\prime)E(z^\prime)}
\label{eq:ap2}
\end{eqnarray}
where $t_0$ is the time denoting the present epoch. 
The function $E(z)$ as obtained from analysis of JLA data \cite{wang:jcap}
is used in Eq. \eqref{eq:ap2}, to obtain $t$ as a
function of $z$ by performing the integration numerically.
We then eliminate $z$ from the obtained $z$ - $t(z)$ dependence and the
 equation $a_0 /a = 1 + z$ to obtain scale factor $a$ as a function
of $t$. 
We then use the time dependence,
$a(t)$, thus obtained, in  Eq.\ (\ref{eq:ome1})
to obtain temporal behaviour of the function $\omega(t)$.
To depict temporal behaviour 
we introduce a dimensionless
time parameter $\tau$ as
\begin{eqnarray}
\tau = \ln a(t)\,.
\label{eq:tau}
\end{eqnarray}
$\tau=0$ corresponds to present epoch 
as the scale factor $a$ is normalised to unity at present epoch.
The cosmic time domain that gets probed in   Supernova Ia observations
is $-0.7 < \tau <0$. 
 In Fig.\  \ref{fig:1}  we have shown
 dependence of the equation of state $\omega$ on time parameter $\tau$ as extracted from the SNe Ia data.
The value $\omega=-1/3$ 
has been depicted by a horizontal line, which corresponds to $\ddot{a} = 0$.
From the
observational data the value $\omega=-1/3$ is realised around $\tau \sim 0.5$ when the universe has undergone a 
transition from the phase of decelerated expansion to accelerated expansion. 
We find that the obtained dependence depicted in Fig.\  \ref{fig:1}  may be best-fitted with a 
polynomial of the form
\begin{eqnarray}
\omega(\tau) = -1 + \sum_{i=0} \beta_i \tau^i
\label{eq:fitomega}
\end{eqnarray}
with coefficients $\beta_i$'s given in table~\ref{tab:1}.
\begin{table}[h]
\begin{center}
\begin{tabular}{|cc|cc|cc|}
\hline
$\beta_0 = $ & -0.704 (-0.631) & $\beta_3 =$ & -2.29 (-3.76) &&\\
\cline{1-4}
$\beta_1 = $ & -0.61 (-0.715)  & $\beta_4 =$ & -2.81 (-4.84) & $\beta_i = 0$&\\ 
\cline{1-4}
$\beta_2 = $ & -0.49 (-1.04)  & $\beta_5 =$ & -0.92 (-1.93) & for $i>5$ &\\
\hline
\end{tabular}
\end{center}
\caption{\label{tab:1}Values of coefficients $\beta_i$'s  giving
best fitting of $\left( -1 +  \sum \beta_i\tau^i\right)$ with $\omega(\tau)$  extracted from
observational data corresponding to $z_{\rm cut}=0$ ($z_{\rm cut}=0.6$).}
\end{table}
In terms of dimensionless time parameter $\tau$, 
the continuity equation
(Eq.\ (\ref{eq:3}))
for the total dark fluid  takes the form
\begin{eqnarray}
\done{}{\tau} \ln(\rho_{\rm de} + \rho_{\rm dm}) &=& -3 -3\omega\,.
\label{eq:17}
\end{eqnarray}

\begin{figure}[t]
\begin{center}
\includegraphics[scale=0.6]{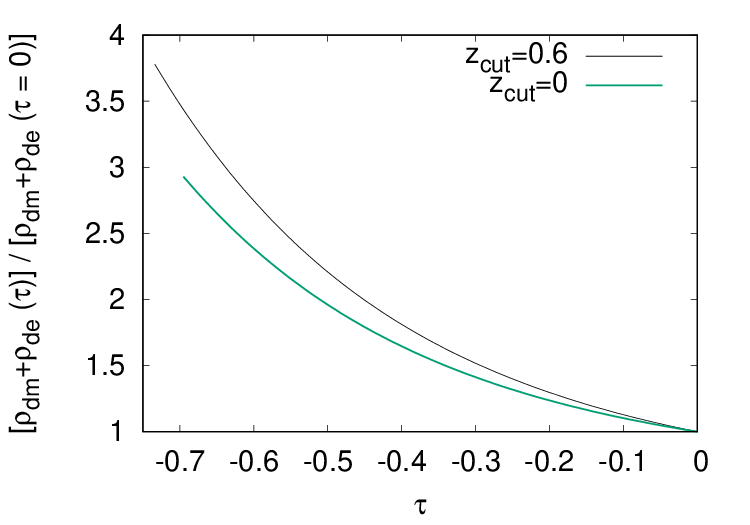} 
\end{center}
\caption{\label{fig:2}Plot of $\frac{[\rho_{\rm de} + \rho_{\rm dm} ]_\tau
}{ [\rho_{\rm de} + \rho_{\rm dm} ]_0}$ vs $\tau$ as obtained from analysis of observational data
for $z_{\rm cut} = 0$ and $z_{\rm cut} = 0.6$ }
\end{figure}

Using the obtained temporal behaviour of  $\omega(\tau)$ 
as shown on Fig.\ \ref{fig:1}  we
solve the above equation performing numerical integration
to obtain $(\tau)$-dependence of  total energy density of
the dark fluid:
\begin{eqnarray}
\Big{[}\rho_{\rm de} + \rho_{\rm dm} \Big{]}_\tau
&=&
{\Big{[}\rho_{\rm de} + \rho_{\rm dm} \Big{]}_0}
\exp\left[-3\int^{\tau}_{\tau^\prime=0}(1+\omega(\tau^\prime)) 
d\tau^\prime \right] \nonumber\\
\label{eq:18}
\end{eqnarray}
The obtained temporal behaviour of total energy density of the dark
fluid is shown in Fig.\ \ref{fig:2}.
We find that the time dependence as depicted in Fig.\ \ref{fig:2}  may be fitted with a polynomial of
the form
\begin{eqnarray}
 [\rho_{\rm de} + \rho_{\rm dm}] &=& \big{[}\rho_{\rm de} + \rho_{\rm dm}\big{]}_0 \sum_{i=0} C_i \tau^i
\label{eq:rhofit}
\end{eqnarray}
\begin{table}[!h]
\begin{center}
\begin{tabular}{|cc|cc|cc|}
\hline
$C_0 = $ & 1 & $C_3 =$ &  -0.65 (-0.62)  && \\
\cline{1-4}
$C_1 = $ & -0.89 (-1.10) & $C_4 =$ & 1.36 (2.096) & $C_i=0$ &\\ 
\cline{1-4}
$C_2 = $ & 1.28 (1.65)  & $C_5 =$ & -0.97 (-1.05) & for $i>5$ &\\
\hline
\end{tabular}
\end{center}
\caption{\label{tab:2}Values of coefficients $C_i$'s  giving
best fitting of polynomial $\sum C_i\tau^i$ with ($\rho_{\rm de}(\tau) + \rho_{\rm dm}(\tau)/ 
[\rho_{\rm de} + \rho_{\rm dm}]_0$) extracted from  observational data corresponding to $z_{\rm cut}=0$ ($z_{\rm cut}=0.6$).}
\end{table}
with coefficients ($C_i$'s) given in table~\ref{tab:2}.

\section{Connection with observation}
\label{sec:4}
In terms of dimensionless parameter $\tau$ introduced in Eq. \eqref{eq:tau},
the non-conservation equation Eq. \eqref{eq:12int} for `fluid 1' component
takes the form 
\begin{eqnarray}
\frac{d}{d\tau} (\ln \rho_1) + 3(1+\omega(\tau)) &=& -  \frac{d\ln V}{d\tau}
\label{eq:rho1tau}
\end{eqnarray}
where we used  $V_0 G = \rho_1$ and the result $d\tau/dt = d(\ln a)/dt = \dot{a}/a = H$.
We choose to parametrise  $\tau$-dependence of $\rho_1$ as
\begin{eqnarray}
\frac{\rho_1 (\tau)}{[\rho_{\rm de} + \rho_{\rm dm} ]_0} & = &  \sum_{i=0}^\infty\alpha_{i}\tau^i   \label{eq:vfit}
\end{eqnarray}
and restrict ourselves to models for which 
\begin{eqnarray}
\frac{d\ln V(\tau)}{d\tau} &=& K_0 \label{eq:vfit1}
\end{eqnarray}
where $K_0$ is independent of $\tau$.
Equation \eqref{eq:vfit1} implies special
class of $k-$essence potentials with temporal behaviour
as $V \sim e^{K_0 \tau} = e^{K_0 \ln a} = [a(t)]^{K_0}$.
For this kind of temporal behaviour of potential, we use Eqs. \eqref{eq:fitomega},\eqref{eq:vfit},\eqref{eq:vfit1} in \eqref{eq:rho1tau} 
 we get,
\begin{eqnarray}
\sum_{i=0}^\infty i \alpha_i \tau^{i-1} +3\sum_{i=0}^\infty \alpha_{i}\tau^i \sum_{j=0}^5 \beta_{j}\tau^j  &=& - K_0\sum_{i=0}^\infty \alpha_{i}\tau^i
\end{eqnarray}
Coefficients of $\tau^i$ in the first and second terms of left hand side
are  $(i+1)\alpha_{i+1}$ and $3\displaystyle\sum_{n=0}^i \beta_n \alpha_{i-n}$ respectively and that in the right hand side is $-K_0\alpha_i$.
Equating these coefficients from both sides we obtain
\begin{eqnarray}
\alpha_{i+1}=  -\frac{(K_0\alpha_i)}{(i+1)}  - 3\sum_{n=0}^i \frac{\beta_n \alpha_{i-n}}{(i+1)} 
\label{eq:recur1}
\end{eqnarray}

\section{Results and discussions}
\label{sec:5}
\begin{figure}[!ht]
\begin{center}
\includegraphics[scale=.6]{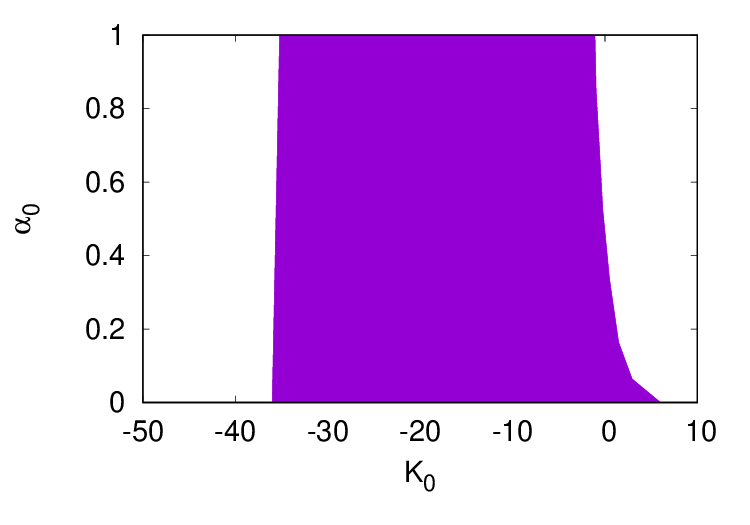} 
\end{center}
\caption{\label{fig:3} Region of  $\alpha_0 - K_0$ parameter space for which 
$\rho_1 > 0$ and $\rho_1 < \rho_{\rm de} + \rho_{\rm dm}$ at all times }
\end{figure}
For a given $K_0$ and $\alpha_0$ ($0<\alpha_0<1$), we may use  
the recursion relation (Eq.\ (\ref{eq:recur1})) to find the
$\alpha_i$'s ($i>1$). Because of the term $(i+1)$ in the denominator
of in Eq.\ (\ref{eq:recur1}), for any given finite $K_0$,
$\alpha_i \approx 0$ for large  $i \gg 1$. For a given
$\alpha_0$ and $K_0$ one may therefore use values of $\alpha_i$
in Eq.\ (\ref{eq:vfit}) to obtain temporal behaviour of
$\rho_1(\tau; \alpha_0,K_0)$. Since $\rho_1$ represents 
the energy density of fluid 1, it must be positive at all time.
Also as $\rho_1 < (\rho_1 + \rho_2) =  (\rho_{\rm de} + \rho_{\rm dm})$,
the condition $0< \rho_1 < (\rho_{\rm de} + \rho_{\rm dm})$
must be  satisfied at all instants of time. 
We use this   to constrain the region of parameter space 
$\alpha_0 - K_0$.\\

The  shaded region in Fig.\ \ref{fig:3} depicts  the allowed 
domain in 
$\alpha_0 - K_0$ parameter space for which 
$0< \rho_1(\tau; \alpha_0,K_0) < (\rho_{\rm de} + \rho_{\rm dm})$ 
at all time.
The results are obtained for the analysis 
of observational data corresponding to $z_{\rm cut} = 0$.
This result shows that 
the allowed spread of $K_0$ is not much sensitive 
to values of $\alpha_0$.
For example, the obtained allowed range of $K_0$ is  
\begin{eqnarray}
&& \mbox{for } \alpha_0 = 0, \quad  -36 < K_0 < 6 \nonumber\\
&& \mbox{for } \alpha_0 = 1, \quad -35 < K_0 < -1 \,.
\label{eq:range}
\end{eqnarray}
Note that the observed data allows negative values   of
$K_0$. As $K_0 = \frac{d\ln V}{d\tau} = \frac{\dot{V}/V}{H}$, 
negative values of $\dot{V}/V$ is  preferred from observation ($H$ being
always positive). For $\dot{V}/V<0$ the term $- G\frac{V_0\dot{V}}{V}$ in Eq.\ (\ref{eq:12int})
representing energy transfer between fluid components `1' and `2' must be positive
as the term $GV_0$ representing energy density of fluid `1' is always positive.
This implies a positive energy transfer from fluid `2'
to fluid `1'.

From Eq.\ (\ref{eq:vfit}) we note that 
\begin{eqnarray}
\alpha_0 = \frac{(\rho_{1})_0}
 {(\rho_{\rm de} + \rho_{\rm dm})_0}
 = \frac{(\rho_{1})_0}
 {(\rho_1  +  \rho_2)_0}
\end{eqnarray}
 where 
 $(\rho_{1})_0$ and $(\rho_{2})_0$ respectively correspond  
 to the value of $\rho_1$ 
 and $\rho_2$  at 
 present epoch ($\tau=0$). This gives
 $ (\rho_1)_0 / (\rho_2)_0  =  \alpha_0 / (1 - \alpha_0)$.
\begin{figure}[t]
\begin{center}
\includegraphics[height=2in, width=3in]{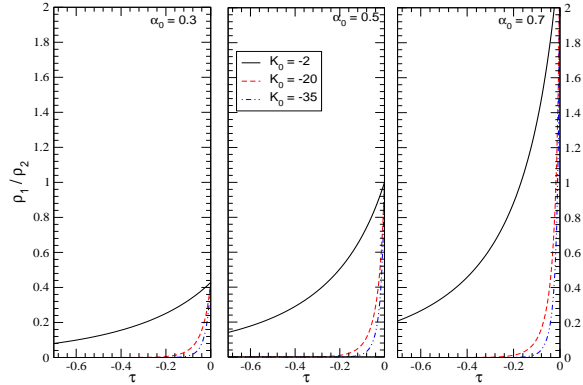} 
\end{center}
\caption{\label{fig:4}Plot of $\rho_1/\rho_2$ vs $\tau$ for different benchmark 
values of $\alpha_0$ and $K_0$ within the allowed range.}
\end{figure}
In Fig.\ (\ref{fig:4})  we have shown temporal behaviour of the 
 ratio $\rho_1/\rho_2$ for different benchmark 
values of $\alpha_0$ and $K_0$ chosen from their allowed range.
Increase in $\rho_1/\rho_2$ with time $\tau$ implies that,
there is more and more transfer of energy from fluid `2'
to fluid `1' as the universe evolves.\\

Finally, the implications of
the result may be stated in the following way.
A $k-$essence model described by Lagrangian $L = V(\phi)F(X)$
with temporal behaviour of $V(\phi(t))$  as $\sim [a(t)]^{K_0}$
with $K_0$ as given in Eq. \eqref{eq:range}
may be used to realise DM-DE interactions in terms of 
diffusive interaction between two ideal fluids `1' and `2'
having equations of 
state same as that of total dark sector of universe.\\

\section{Conclusion}
\label{sec:6}
In this work we have 
exploited   dynamics
of a $k-$ essence scalar field governed by
the Lagrangian $L = V(\phi) F(X)$ to show that 
the interaction between
dark energy and dark matter component of the universe
may be realised in terms of diffusive interaction
between two ideal fluids `1' and `2'. 
These two fluids may respectively 
be represented by two $k-$essence
Lagrangians of  the form: $L_1 = V_0 F(X)$
and $L_2 = V_1(\phi) F(X)$, where $V(\phi) = V_0 + V_1(\phi)$. 
Such a realisation is also shown to be 
consistent with observed data 
on late time cosmic acceleration,
for some constraints on temporal behaviour of $k-$essence potential $V(\phi)$.
We have discussed in the paper the methodology of obtaining 
such constraints from analysis of SNe Ia data. 
In particular, we have found that
for a special class of $k-$essence potentials having temporal behaviour as $ V(\phi(t)) \sim [a(t)]^{K_0}$
where $K_0$ is a constant and $a(t)$ is the scale factor, the value of $K_0$ is constrained within
the range $-36 < K_0 < 6$.

\section*{Acknowledgments}
The authors are grateful to the referee for very helpful suggestions.
A.C. would like to thank University Grants Commission (UGC), India
for supporting this work by means of NET Fellowship (Ref.No.22/06/2014(i)EU-V
and Sr.No.2061451168). 
%

\end{document}